\documentclass[fleqn,10pt]{wlscirep}

\usepackage{amssymb}
\usepackage{amsmath}
\usepackage{mathrsfs}
\usepackage{bbold}
\usepackage{graphicx}
\usepackage{psfrag}
\usepackage{multirow}
\usepackage{hyperref}
\usepackage{xcolor}

\usepackage[justification=justified]{caption}

\usepackage{tikz}
\usetikzlibrary{matrix,positioning}
\usetikzlibrary{arrows}
\usetikzlibrary{decorations.markings}
\usetikzlibrary{shapes.geometric}

\newcommand{\be}{\begin{equation}}
\newcommand{\ee}{\end{equation}}

\newcommand{\phase}[1]{\mathbb{#1}}
\renewcommand{\neg}[1]{\overline{#1}}

\makeatletter
\def\Ddots{\mathinner{\mkern1mu\raise\p@
\vbox{\kern7\p@\hbox{.}}\mkern2mu
\raise4\p@\hbox{.}\mkern2mu\raise7\p@\hbox{.}\mkern1mu}}
\makeatother

\begin{document}

\title
{The exact phase diagram for a class of open multispecies asymmetric exclusion processes}
\author[1,*]{Arvind Ayyer}
\author[1]{Dipankar Roy}

\affil[1]{Department of Mathematics, 
Indian Institute of Science,
Bangalore - 560012, India}

\affil[*]{arvind@iisc.ac.in}

\date{\today}


\begin{abstract}
The asymmetric exclusion process is an idealised stochastic model of transport, whose exact solution has given important insight into a general theory of nonequilibrium statistical physics. In this work, we consider a totally asymmetric exclusion process with multiple species of particles 
on a one-dimensional lattice in contact with reservoirs. We derive the exact nonequilibrium phase diagram for the system in the long time limit.
We find two new phenomena in certain regions of the phase diagram: {\em dynamical expulsion} when the density of a species becomes zero throughout the system, and {\em dynamical localisation} when the density of a species is nonzero only within an interval far from the boundaries. 
We give a complete explanation of the macroscopic features of the phase diagram using what we call {\em nested fat shocks}. 
\end{abstract}

\flushbottom
\maketitle

\section*{Introduction}
The one-dimensional asymmetric simple exclusion process (ASEP) with open boundaries has been of great importance as a model system towards understanding nonequilibrium phenomena. The matrix ansatz, which has since become an important tool \cite{be}, was developed to compute the nonequilibrium steady state (NESS) of the totally asymmetric version of the single-species ASEP \cite{dehp} (TASEP). Using this ansatz, various measurable quantities such as the density and current were calculated in the NESS, from which the exact nonequilibrium phase diagram  was derived. While many rigorous results are known in one dimension, there are none for higher dimensional exclusion processes in contact with reservoirs.

The exact calculation of various out-of-equilibrium quantities was also useful in formulating general principles. For example, the diffusion constant was first calculated exactly for the open TASEP \cite{derrida-evans-mallick-1995}. Using the integrability of the ASEP, the spectrum of the transition matrix was computed \cite{degier-essler-2005}. The large deviation functional for the density profiles \cite{derrida-lebowitz-speer-2003} and the large deviation function for the current \cite{degier-essler-2011, lazarescu-mallick-2011} was derived using the matrix ansatz.
These calculations helped in the formulation of two general principles for driven diffusive systems; the additivity principle \cite{bodineau-derrida-2004} and the macroscopic fluctuation theory \cite{BDJGL-2005}.

It is natural to extend the ASEP to several species of particles. Multispecies exclusion processes have applications to studies of traffic flow~\cite{karimipour-1999}, cell motility \cite{simpson-et-al-2009, penington-et-al-2011},  chemotaxis \cite{painter-2009}, chemical reactions \cite{bruna-chapman-2012} and biological transport in ion-channels~\cite{agmon-1995,zilman-et-al-2007, jovanovic-et-al-2009}.
The ASEP with second-class particles  was first considered with periodic boundary conditions \cite{djls}. The first model with two species and open boundaries whose NESS was determined exactly was a model where positive and negative charges moved in the lattice under the influence of an electric field~\cite{evans-et-al-1995}. Much later, a two-species model called the  {\em semipermeable TASEP} \cite{arita-2006,als-2009} was considered, where second-class particles were confined to the lattice.
The computation of the NESS was later generalised to the semipermeable ASEP \cite{uchiyama-2008}.
The NESS of a version with more general boundary conditions was also determined exactly \cite{als-2012}. Later on, the phase diagram for a large class of ASEPs with two species and open boundaries was obtained \cite{CMRV-2015, CEMRV-2016}. More recently, classes of integrable ASEPs with multiple species of particles and open boundaries were determined \cite{crampe-et-al-2016}.

In this work, we study a multispecies exclusion process on a finite one-dimensional lattice with $r$ species of charges called the mASEP introduced recently \cite{cantini-etal-2016}. The hopping rates in the bulk are asymmetric and those in the boundary are defined in such a way that there are $r+1$ conserved particle numbers. 
The main results are the following.  
We obtain the complete $(r+2)$-dimensional phase diagram and present formulas for all densities and currents in the thermodynamic limit in all regions of the phase diagram. 
It will turn out that all the macroscopic features can be explained by a new structure which we call a {\em nested fat shock}. 
To make the presentation self-contained, we review the features for the semipermeable ASEP in Section~I of the Supplementary material. We will prove these results in Section~II of the Supplementary material by using projections to the semipermeable ASEP.

\section*{Results} 
\label{sec:model}

The mASEP is defined on a one-dimensional lattice of size $n$, where each site is occupied by exactly one particle of type $\{\neg{r},\dots,\neg{1},0,1,\dots,r\}$.
The barred particles are negative charges, the unbarred ones are positive charges, and 0's are vacancies. There are $r$ species of charges, with the total number of particles of charge $j$ being fixed to be $n_j$ for  $1\leq j\leq r$. As a consequence, the number $n_0$ of vacancies is also fixed, with $n_0 + \cdots + n_r = n$. 
More precisely, fix an $(r+1)$-tuple of positive integers $\underline{n} = (n_0,\dots,n_r)$. The state space $\Omega_{\underline{n}}$ consists of all words of length $n$ in the alphabet $\{\neg{r},\dots,\neg{1},0,1,\dots,r\}$ such that the total number of $j$'s and $\neg j$'s is equal to $n_j$ for $1 \leq j \leq r$ and the total number of $0$'s is $n_0$.

\begin{figure}[h]
\begin{center}

\includegraphics[width=\textwidth]{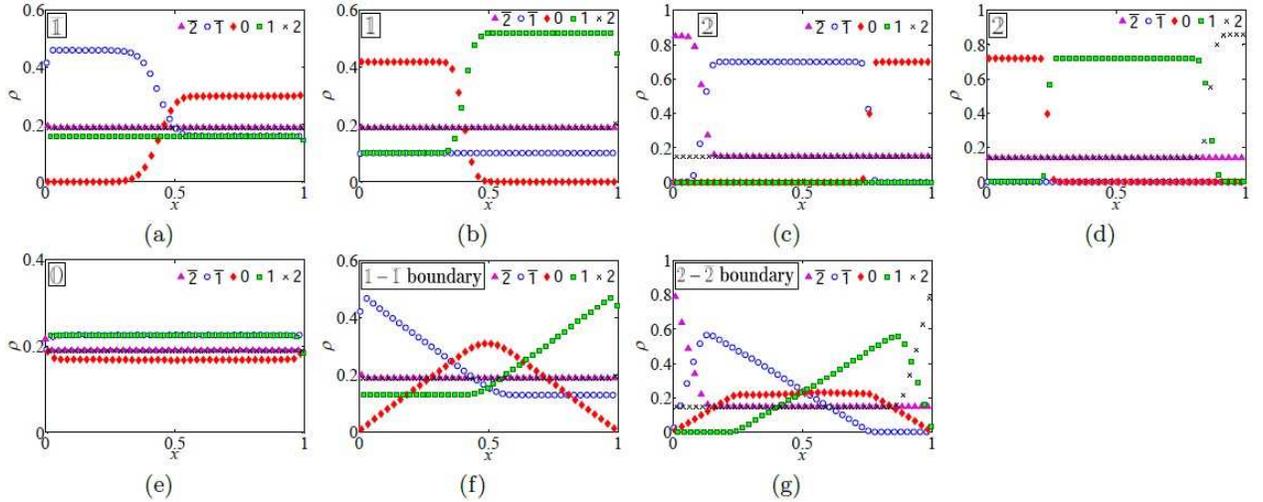}

\caption{Plots of the densities of particles 2 (black crosses), 1 (green squares), 0 (red diamonds), $\neg{1}$ (blue circles), and $\neg{2}$ (pink triangles), versus the scaled position $x=m/n$ for the mTASEP with  $r=2$ where $n=1000$, $\theta_0 =0.17, \theta_1=0.45$ and $\theta_2=0.38$
 in the regions (a) $\phase{\neg{1}}$ ($\alpha=0.35, \beta= 0.83 $), 
 (b) $\phase{1}$ ($\alpha=0.73, \beta=0.29 $),  
 (c) $\phase{\neg{2}}$ ($\alpha=0.15, \beta=0.81 $), 
 (d) $\phase{2}$ ($\alpha=0.73, \beta=0.14 $), 
 (e) $\phase{0}$ ($\alpha=0.71, \beta=0.87 $), 
 (f) the $\phase{1} - \phase{\neg{1}}$ shock line ($\alpha=\beta=0.32  $), and
 (g) the $\phase{2} - \phase{\neg{2}}$ shock line ($\alpha=\beta=0.15$).}
\label{fig:2-mTASEP}
\end{center}
\end{figure}

The dynamics is the effect of a rightward-pointing electric field. In the bulk, we have the asymmetric hopping rule
\be \label{bulk-rule}
i\; j\longrightarrow j\; i\quad\text{with rate}
\begin{cases}
1 & \text{if $i>j$,} \\
$q$ & \text{if $i<j$,}
\end{cases}
\ee
where we think of the barred particles as negative numbers, and set $q<1$. On the left and right boundaries, positive charges can only replace and be replaced by their negatively charged partners with rates given by
\be \label{bound-rule}
\begin{array}{c| c}
\text{\underline{Left}} & \text{\underline{Right}} \\[0.8mm]
\neg{j} \longrightarrow j \quad \text{with rate $\alpha$}, &
\; j \longrightarrow \neg{j} \quad \text{with rate $\beta$},\\
j \longrightarrow \neg{j} \quad \text{with rate $\gamma$}, &
\; \neg{j} \longrightarrow j \quad \text{with rate $\delta$}.
\end{array}
\ee
The mASEP possesses charge-conjugation symmetry in the following sense:  interchanging positively and negatively charged particles as well as the rates $1$ and $q$, $\alpha$ and $\beta$, and $\gamma$ and $\delta$, and changing the direction of motion leaves the mASEP invariant.
The model with $r=1$ and $n_0=0$ is the single-species open ASEP \cite{dehp}; for arbitrary $n_0$, this is the semipermeable ASEP \cite{uchiyama-2008}. Furthermore, if $q=\gamma=\delta=0$, this is the semipermeable TASEP \cite{arita-2006,als-2009} (see Section~I of the Supplementary material). We thus denote the mASEP with $q=\gamma=\delta=0$ as the mTASEP. Results of simulations for the mTASEP with $r=2$ are given in Fig.~\ref{fig:2-mTASEP}.

\subsection*{Phase Diagram}

\begin{figure}[h]
\begin{center}

\includegraphics[width=0.6\textwidth]{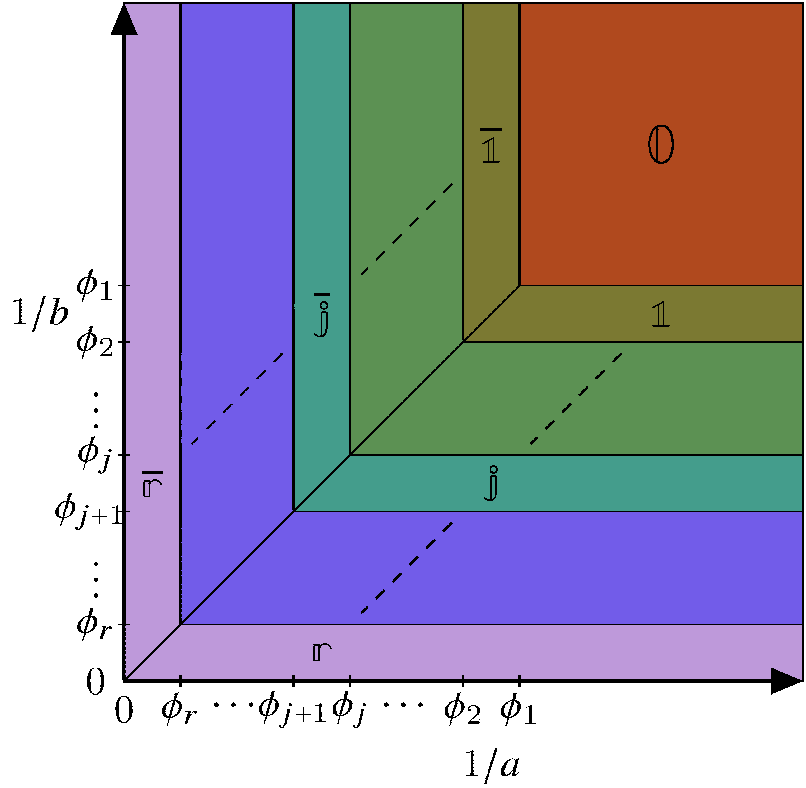}

\caption{The nonequilibrium phase diagram of the mASEP with $r$ species of charges. There are $2r+1$ different regions, which are labelled $\phase 0, \phase 1, \dots, \phase r, \phase{\neg 1}, \dots, \phase{\neg r}$. Each region is characterised by different bulk densities of all particles. The explanation for the nomenclature of the phases in given below.
See Table~\ref{tab:dens} and Table~\ref{tab:curr} for the densities and currents respectively in the NESS in these $2r+1$ regions.}
\label{fig:phase-diag-2r}
\end{center}
\end{figure}

For each integer $j$ between 0 and $r$, let $\theta_j=n_j/n$ be the total density of (both positively and negatively charged) particles of species $j$. We consider the behaviour of the mASEP in the limit $n\to\infty$ and $n_j\to\infty$ for each $j$ such that the total density of species $j$ particles converges to $\theta_j > 0$. Define the quantities
\be
\begin{split}
\label{ab}
a &= \frac{1-q- \alpha+\gamma+\sqrt{(1-q- \alpha+\gamma)^2+4 \alpha \gamma}}
{2 \alpha}, \\
b  &= \frac{1-q- \beta+\delta+\sqrt{(1-q- \beta+\delta)^2+4 \beta \delta}}
{2 \beta},
\end{split}
\ee
and the function $f(x)=1/(1+x)$.
Set $\Theta_k=(\theta_k+\cdots+\theta_{r})/2$
and $\phi_k=\Theta_{k}/(1-\Theta_{k})$ for $1\leq k\leq r$. 
Then the exact phase diagram is given in Fig.~\ref{fig:phase-diag-2r}.

\begin{table}[h]
\begin{center}

\begin{tabular}{|c|c|c|}
\hline
Phase & \multicolumn{2}{|c|}{Densities in the bulk} \\ \cline{2-3}
$\downarrow$ & Species $k$ & Values of $\rho_k,\rho_{\neg{k}}$ \\
\hline
\multirow{2}{*}{$\phase{0}$} 
& $k=0$ & $\rho_0 =\theta_0$ \\ \cline{2-3}
 & $1 \leq k \leq r$ & $\rho_k=\rho_{\neg{k}}=\theta_k/2$ \\
\hline
\multirow{6}{*}{$\phase{j}$} & $k=0$ & $\rho_0$ piecewise constant \\ \cline{2-3}
& \multirow{2}{*}{$1 \leq k \leq j-1$} & $\rho_{\neg{k}}=0$ \\ 
& & $\rho_k$ piecewise constant \\ \cline{2-3}
& \multirow{2}{*}{$k=j$} & $\rho_{\neg{j}}=f(b) - \Theta_{j+1}$ \\ 
& & $\rho_j$ piecewise constant \\ \cline{2-3}
& $j+1 \leq k \leq r$ & $\rho_k=\rho_{\neg{k}}=\theta_k/2$ \\ \cline{2-3}
\hline
\multirow{6}{*}{$\phase{\neg{j}}$} 
& $k=0$ & piecewise constant \\ \cline{2-3}
& \multirow{2}{*}{$1 \leq k \leq j-1$} & $\rho_{k}=0$ \\
& & $\rho_{\neg{k}}$ piecewise constant \\ \cline{2-3}
& \multirow{2}{*}{$k=j$} & $\rho_{j}=f(a) - \Theta_{j+1}$  \\
& & $\rho_{\neg{j}}$ piecewise constant \\ \cline{2-3}
& $j+1 \leq k \leq r$ & $\rho_k=\rho_{\neg{k}}=\theta_k/2$ \\
\hline
\multirow{3}{*}{$\phase{j}-\phase{\neg{j}}$ boundary}
& $k=0$ & $\rho_0$ piecewise linear \\ \cline{2-3}
& $1 \leq k \leq j-1$ & $\rho_k, \rho_{\neg{k}}$ piecewise linear  \\ \cline{2-3}
& $j+1 \leq k \leq r$ & $\rho_k=\rho_{\neg{k}}=\theta_k/2$ \\
\hline
\end{tabular}
\caption{The densities of all species of particles in phase $\phase{0}$, as well as phases $\phase{j}$ and $\phase{\neg{j}}$, and the $\phase{j}-\phase{\neg{j}}$ boundary for $1\leq j\leq r$. Piecewise constant densities correspond to phase separation and piecewise linear profiles correspond to averaging over shocks. The exact formulas can be calculated from the schematic plots in the top row of Fig.~\ref{fig:nested fat shock}.}
\label{tab:dens}

\end{center}
\end{table}

The nonequilibrium phase diagram of the $r$-species mASEP in Fig.~\ref{fig:phase-diag-2r} comprises $2r+1$ phases, 
$\phase{\neg{r}},\dots,\phase{\neg{1}},\phase{0},\phase{1},\dots,\phase{r}$,
each one of which is characterised by the bulk densities $\rho_j$ for $j \in \{\neg r, \dots, 0, \dots, r \}$,
as well as the currents $J_j$ for $j \in \{1,\dots,r\}$ of all types of particles; we tabulate these in Table~\ref{tab:dens} and \ref{tab:curr} respectively. 
Note that there is no current of $0$'s and the current of $\neg j$'s to the left is the same as that of $j$'s to the right.
The mean densities of the $j$ and $\neg{j}$ jump discontinuously across the $\phase{j}-\phase{\neg{j}}$ boundary. By contrast, the mean densities vary continuously along the $\phase{j}-(\phase{j+1})$ (and $\phase{\neg{j}}-(\phase{\neg{j+1}})$) boundary. All currents $J_j$ change continuously across all phase boundaries in  Fig.~\ref{fig:phase-diag-2r}. In all phases except $\phase{0}$, the system shows phase coexistence with a sharp interface separating intervals of different density for some particle type, as we show by the illustrative density profiles in Fig.~\ref{fig:2-mTASEP} for the mTASEP with $r=2$ and Supplementary Fig. S2 for the mASEP with $r=1$.
The proofs of the density profiles in Table~\ref{tab:dens} and currents in Table~\ref{tab:curr}, which explain the phase diagram in Fig.~\ref{fig:phase-diag-2r}, are given in Section~II of the Supplementary material.

\begin{figure}[h]
\begin{center}

\includegraphics[scale=1]{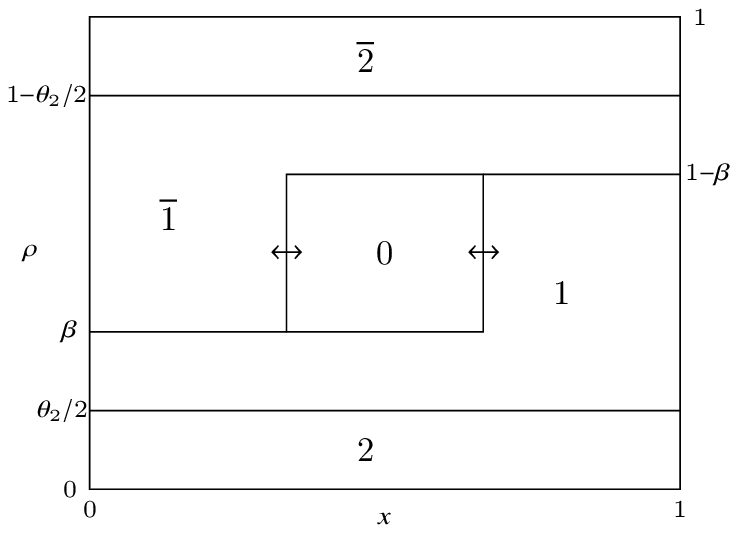}

\caption{Instantaneous picture of the nested fat shock in the $\phase{1}-\phase{\neg{1}}$ boundary in the 
rescaled mTASEP with $r=2$.
Each connected region is labelled with the species of a particle and the height of a region at a given
location represents the density of that species at that point.}
\label{fig:nested fat shock-eg1}

\end{center}
\end{figure}

\begin{figure}[h]
\begin{center}

\includegraphics[scale=1]{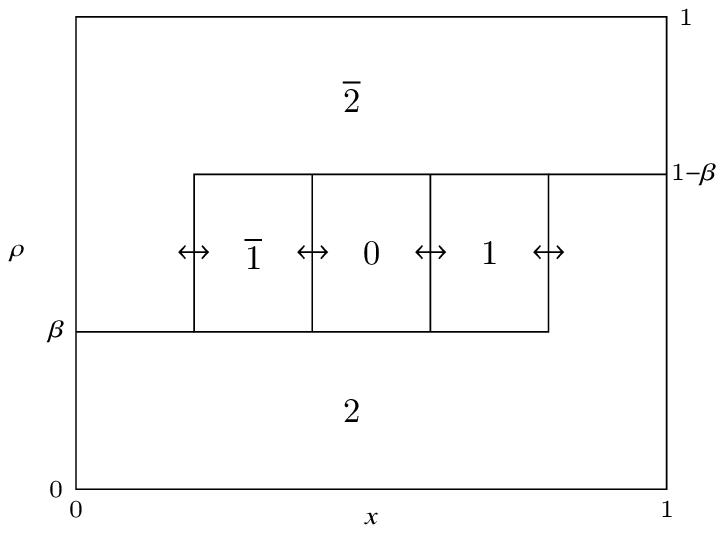}
\caption{Instantaneous picture of the nested fat shock in the $\phase{2}-\phase{\neg{2}}$ boundary
in the rescaled mTASEP with $r=2$.
Each connected region is labelled with the species of a particle and the height of a region at a given
location represents the density of that species at that point.}
\label{fig:nested fat shock-eg2}
\end{center}
\end{figure}

We illustrate the density profiles for various regions in the phase diagram for the example of the $2$-species mTASEP 
(i.e. $q = \gamma = \delta = 0$) in Fig.~\ref{fig:2-mTASEP}. Therefore, $a = (1-\alpha)/\alpha$ and $b=(1-\beta)/\beta$ from \eqref{ab}, 
and $f(a) = \alpha$, $f(b) = \beta$. There are five phases for this system, corresponding to $r=2$, and two relevant phase-boundaries. 
The currents can be calculated by mean-field type calculations from the densities in each of the regions. More precisely, 
$J_2 = \rho_2 (1-\rho_2) = \rho_{\neg 2} (1-\rho_{\neg 2})$ and $J_1 = \rho_1 (1-\rho_1 - 2\rho_2) 
= \rho_{\neg 1} (1-\rho_{\neg 1} - 2\rho_{\neg 2})$. When there is phase-separation, the densities of all species conspire to ensure that the
currents are constant across the system (because of particle conservation in the bulk). The value of the currents in each phase can be compared with Table~\ref{tab:curr}.

In phase $\phase 0$, all densities are constant, and the densities of oppositely charged particles are equal. Therefore, $\rho_0 = \theta_0$, and $\rho_k = \rho_{\bar k} = \theta_k/2$ for $k=1,2$.
This is seen in Fig.~\ref{fig:2-mTASEP}(e) and matches with the first row of Table~\ref{tab:dens}. 

In phase $\phase 1$, the densities of $2$'s, $\neg 2$ and $\neg 1$'s are constant, whereas those of $1$'s and $0$'s undergo phase separation. 
As in phase $\phase 0$, $\rho_2 = \rho_{\bar 2} = \theta_2/2$. Moreover, $\rho_{\bar 1} = \beta - \theta_2/2$ is also constant throughout the system. In the phase-separated regions, either $\rho_1 + \rho_{\bar 1} = 1-\theta_2$ (forcing $\rho_0 = 0$), or $\rho_1 = \rho_{\bar 1}$ from which $\rho_0$ can be calculated. The density plots can be seen in Fig.~\ref{fig:2-mTASEP}(b) and match the calculation of the densities in the second row of Table~\ref{tab:dens} with 
$\phase j  = \phase 1$. The density profiles in $\phase{\neg 1}$ can be calculated analogously using charge-conjugation symmetry.

In phase $\phase 2$, the only constant densities are given by $\rho_{\bar 2} = \beta$ and $\rho_{\bar 1} = 0$. It is not immediately obvious why 
$\bar 1$'s are excluded from the system, and we give an explanation for this phenomenon of {\em dynamical expulsion} in the next section. Particles of species $0,1$ and $2$ are phase segregated in three distinct parts. The density of $2$'s, $\rho_2$, is given by the $1-\beta$ in the rightmost part and $\rho_{\bar 2}$ in the other two parts. Particles of species $0$ exist only in the leftmost part with density $1-2\beta$, and those of species $1$ 
exist only in the middle part with the same density. In the thermodynamic limit, the middle part is infinitely far away from both boundaries and it is not immediately clear how $1$'s can be localised in the bulk. We call this phenomenon {\em dynamical localisation} and explain how this occurs in the next section. The densities can be seen in Fig.~\ref{fig:2-mTASEP}(d) and match the calculation of the densities in the second row of Table~\ref{tab:dens} with $\phase j  = \phase 2$. Again, the profiles in $\phase{\neg 2}$ can be calculated using charge-conjugation symmetry.

The nomenclature for the phases can now be explained. Each phase is denoted by the phase-segregated species with largest absolute value. For example, $0$'s and $1$'s are segregated in phase $\phase 1$, $0$'s and $\neg 1$'s are segregated in phase $\phase{\neg 1}$, 
$0$'s, $1$'s and $2$'s are segregated in phase $\phase 2$, and $0$'s, $\neg 1$'s and $\neg 2$'s are segregated in phase $\phase 2$. The sole exception is phase $\phase 0$, where all species have constant density.

To understand the density profiles in the $\phase{1}-\phase{\neg{1}}$ and $\phase{2}-\phase{\neg{2}}$ boundaries, we appeal to the nested fat shock construction, which we explain below. Recall that the phase diagram is calculated in the limit where the system size, $n \to \infty$. We rescale the system by a factor of $1/n$ so that the locations lie in the interval $[0,1]$.
In the $\phase{1}-\phase{\neg{1}}$ boundary, as shown in Fig.~\ref{fig:nested fat shock-eg1},
the densities of particles $2$ and $\neg 2$ are constant and equal to $\theta_2/2$. All the particles of type $0$ form a `bound state' of fixed width. 
We call this the nested fat shock (the nesting is of order 1 here).
Both ends of the bound state execute a synchronised symmetric random walk with reflecting boundary conditions. 
As a consequence, $\rho_{\neg 1}$, $\rho_0$ and $\rho_1$ are piecewise linear after averaging.
In particular, $\rho_1$ is constant towards the left, since the right end of the bound state cannot move all the way to the left, and similarly for $\rho_{\neg 1}$. This is shown in Fig.~\ref{fig:2-mTASEP}(f). 

In the $\phase{2}-\phase{\neg{2}}$ boundary as shown in Fig.~\ref{fig:nested fat shock-eg2}, none of the densities are constant, and the picture is more complicated. The nested fat shock here consists of the regions containing $\neg 1$'s, $0$'s and $1$'s, in that order from left to right. 
The nesting is of order 2 here. There are four boundaries between the regions $\neg 2-\neg 1$, $\neg 1-0$, $0-1$ and $1-2$, and all of them perform synchronised symmetric random walks in the bulk so that the widths of the regions containing $\neg 1$, $0$ and $1$ is fixed. 
When one of them touches the boundary the widths of either $\neg 1$ or $1$ can decrease, causing the opposites charged region to increase in size so that the sum of the widths of these two remains constant. The width of the region containing $0$ never changes.
This behaviour results in the piecewise linear profile shown in Fig.~\ref{fig:2-mTASEP}(g).

One can now derive the density profiles in regions $\phase 1$, $\phase 2$, $\phase{\neg 1}$ and $\phase{\neg 2}$ from these nested fat shocks. For example, in $\phase 1$, one has the same nested fat shock structure as in Fig.~\ref{fig:nested fat shock-eg1}, but the ends of the bound state containing $0$ execute a random walk with negative drift, which ensures that the nested fat shock is pinned to the left. Similarly, the nested fat shock is pinned to the right in $\phase{\neg 1}$. Similarly, the density profiles in $\phase 2$ and $\phase{\neg 2}$ can be calculated by forcing the nested fat shock in Fig.~\ref{fig:nested fat shock-eg2} to be pinned to the left and right respectively.

The general structure of the nested fat shock is explained in the next section.

\begin{table}[h]
\begin{center}
\begin{tabular}{|c|c|c|}
\hline
Phase & \multicolumn{2}{|c|}{Currents} \\[0.4mm] \cline{2-3}
$\downarrow$ & Species $k$ & Value of $J_k$ \\[0.4mm]
\hline
\multirow{2}{*}{$\phase{0}$} 
& $1 \leq k < r$ & $(1-q) (\Theta_k - \Theta_{k+1}) (1 - \Theta_k - \Theta_{k+1})$ \\[0.4mm] \cline{2-3}
 & $k = r$ & $(1-q)\Theta_r(1-\Theta_r)$ \\[0.4mm]
\hline
\multirow{4}{*}{$\phase{j}$} & $1 \leq k \leq j-1$ & $0$ \\ \cline{2-3}
& $k=j$ & $(1-q) \left( f(b)(1-f(b)) -\Theta_{j+1} (1 - \Theta_{j+1}) \right)$ \\ \cline{2-3}
& $j+1 \leq k < r$ & $(1-q) (\Theta_k - \Theta_{k+1}) (1 - \Theta_k - \Theta_{k+1})$ \\ \cline{2-3}
& $k = r$ & $(1-q)\Theta_r(1-\Theta_r)$ \\ 
\hline
\multirow{4}{*}{$\phase{\neg j}$} & $1 \leq k \leq j-1$ & $0$ \\ \cline{2-3}
& $k=j$ & $(1-q) \left( f(a)(1-f(a)) -\Theta_{j+1} (1 - \Theta_{j+1}) \right)$ \\ \cline{2-3}
& $j+1 \leq k < r$ & $(1-q) (\Theta_k - \Theta_{k+1}) (1 - \Theta_k - \Theta_{k+1})$ \\ \cline{2-3}
& $k = r$ & $(1-q)\Theta_r(1-\Theta_r)$ \\ 
\hline
\end{tabular}

\caption{The currents of all species of particles in phase $\phase{0}$, as well as phases $\phase{j}$ and $\phase{\neg{j}}$ for $1\leq j\leq r$. All currents are seen to be continuous across the $\phase{j}-\phase{\neg{j}}$ boundary. For the special cases of $\phase{r}$ and $\phase{\neg{r}}$, take $\Theta_{r+1} = 0$.
Note that $J_0 = 0$ and $J_{\neg k} \equiv -J_k$.}
\label{tab:curr}
\end{center}
\end{table}

\subsection*{Nested fat shock}
\label{sec:nested}

All coarse features of the phase diagram in Fig.~\ref{fig:phase-diag-2r} are explained by the nested fat shock construction.
This is a generalisation of the {\em fat shock} construction, which explains the phase diagram for the semipermeable TASEP~\cite{als-2009} (i.e. the mTASEP with $r=1$). Roughly speaking, the fat shock consists of a macroscopic interval of the system, where all the $0$'s are localised. The $0$'s form two simultaneous shocks with the $1$'s and $\neg 1$'s, with a constant macroscopic width. For more details on the fat shock, see Section~I of the Supplementary material.

The nested fat shock is a macroscopic interval of the system where, for some $j \geq 0$, particles of species $\neg{j},\dots,0,\dots,j$ are localised in a 
very specific way. 

Particles of species $0$  have a nonzero constant density in a subinterval of fixed width inside this interval. Particles of species $1$ (resp. $\neg 1$) have a nonzero constant density in a subinterval to the right (resp. left) of the $0$'s. Although the widths of the $1$ and $\neg 1$ subintervals may vary, the sum of their widths is fixed. This pattern continues until species $j$ on the right and species $\neg j$ on the left. The boundary between any two adjacent subintervals is a shock-front. 
Depending on which part of the phase diagram the system finds itself in, the nested fat shock can have either negative, positive or zero drift. If the drift is negative, the negatively charged subintervals containing $\neg 1, \dots, \neg j$ will not exist, and similarly if there is positive drift, the positively charged subintervals $1,\dots,j$ will vanish. If there is zero drift, all subintervals will exist and move in a synchronised fashion.

\begin{figure}[h]
\begin{center}

\includegraphics[width=\textwidth]{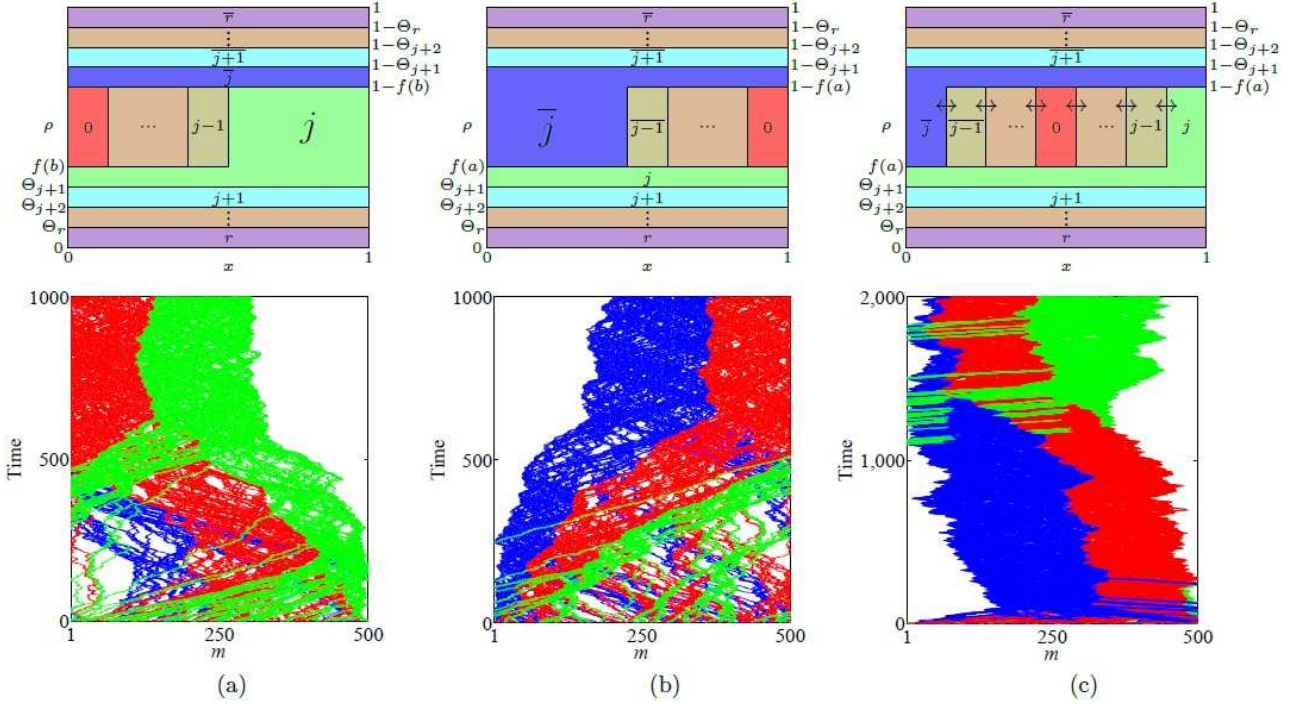}

\caption{
The top row shows schematic plots of the densities $\rho_j$ and
$\rho_{\neg{j}}$, for all $j$, versus  the normalised position $x$ illustrating
the nested fat shock (a) pinned to the left in region $\phase{j}$, (b) pinned to the right in region $\phase{\neg j}$ and (c) unpinned in the $\phase{j}-\phase{\neg j}$ boundary in (c). The densities $\rho_a(x)$ are plotted against the rescaled location $x$. 
The value of $\rho_a(x)$ is the height of the region containing particle $a$ at $x$. 
The bottom row shows simulation results in multiples of 2000 steps as spatiotemporal plots for the mTASEP with $r=2$ and $n_0 = 70, n_1 = 100, n_2 = 330$ in (a) region~$\phase{2}$ ($\alpha=0.79, \beta=0.23$), (b) region~$\phase{\neg 2}$ ($\alpha=0.25, \beta=0.73$), and (c) the $\phase{2}-\phase{\neg 2}$ boundary ($\alpha= \beta=0.28$). The blue, red and green colours represent $\neg 1, 0$ and $1$ particles respectively. The particles of type $2$ and $\neg 2$ are shown in white.
See the hyperlinked videos \href{https://drive.google.com/open?id=0B93XrncWWJUtb05WNExMbW94T00}{here}, \href{https://drive.google.com/open?id=0B93XrncWWJUtNzdreEhLeHZnX00}{here} and \href{https://drive.google.com/open?id=0B93XrncWWJUtbWNYWXNDbnl1bWs}{here} showing the spatiotemporal evolution of the nested fat shocks in simulation runs for (a), (b) and (c) respectively.
}
\label{fig:nested fat shock}

\end{center}
\end{figure}

The top row of Fig.~\ref{fig:nested fat shock} shows the structure of the nested fat shock in these three cases in the most general scenario.
We give a concrete example of a simulation run of the mTASEP  with $r=2$ in the bottom row of the figure,
which shows the results for phases 
(a) $\phase 2$ (negative drift), (b) $\phase{\neg 2}$ (positive drift), and (c) the $\phase 2-\phase{\neg 2}$ boundary (zero drift).
The simulations show the densities of particles of species $\neg 1, 0$ and $1$ only. 
If the nested fat shock has nonzero drift, it gets pinned to one of the boundaries; it is pinned on the left in (a) and on the right in (b).
When the nested fat shock is pinned to the left, it consists of species $0$ and $1$, and particles of species $\neg 1$ exit the system from the left. Similarly, when pinned to the right, it consists of species $\neg 1$ and $0$, and particles of species $1$ vacate the system from the right.
When the nested fat shock has zero drift as in (c), blocks of $\neg 1$, $0$ and $1$ are present, with all shock fronts performing lockstep symmetric random walks.

The precise details for each phase are given below. 

\subsubsection*{Region $\phase 0$}
\label{sec:reg-unif}

Here, only the particles of species $0$ participate in the nested fat shock and the width of the shock is larger than the system size. As a result, all densities are constant throughout the system.
This explains the densities and currents in region $\phase 0$.
See the density plot in Fig.~\ref{fig:2-mTASEP}(e) for the result of simulations.

\subsubsection*{Regions $\phase j$ and $\phase{\neg j}$}
\label{sec:dyn-loc-ej}

We give details only for region~$\phase j$, since those of $\phase{\neg j}$ can be obtained by analogous arguments.

In region~$\phase{j}$, particles of species $\neg{j-1},\dots,0,\dots,j-1$ participate in the nested fat shock. The velocity of this shock is negative and it gets pinned to the left boundary. However, this is not a stable situation. Initially, particles of species $\neg{j-1}$ will be replaced by those of species $j-1$, which will then move rightwards in the bulk, until they join the subinterval of the nested fat shock occupied by the particles of species $j-1$. Once that process is completed, a similar phenomenon will happen with particles of species $\neg{j-2}$. This process will continue until all negatively charged particles in the nested fat shock have been replaced by their positive counterparts. In 
the steady state, we will only see particles  of species $0,\dots,j-1$ participating in the shock, which will be pinned to the left. 

We point out two new nonequilibrium features of this phase which can be seen in  the top row of Fig.~\ref{fig:nested fat shock}(a). First, note that species $0$ through $j-1$ are localised in the interior of the system. Each of these species has undergone phase separation, with one region of non-zero density and the remaining of zero density.
The precise locations of the region with non-zero density can be calculated from the values of $\theta_0,\dots,\theta_{j-1}$. What is more interesting is that species $1$ through $j-1$ are localised {\em away from the boundary}. This is somewhat counterintuitive since we have taken the thermodynamic limit and these particles are infinitely far away from the boundary. We call this phenomenon {\em dynamical localisation}. Such a phenomenon cannot occur in an equilibrium system.
The second new feature is the complete absence of particles of species $\neg{1},\dots,\neg{j-1}$ in the system, i.e. $\rho_{\neg 1} = \cdots = \rho_{\neg{j-1}} = 0$. This is related to the previous phenomenon since these particles can only enter at the expense of the dynamically localised particles. We call this phenomenon {\em dynamical expulsion}. In the extreme case of the $\phase r$ phase, all the barred particles except $\neg{r}$ are expelled. 

The picture in the $\phase{\neg j}$ phase can be derived analogously.
See the density plots in Regions $\phase{\neg 1}, \phase 1, \phase{\neg 2}$ and $\phase 2$ of Fig.~\ref{fig:2-mTASEP}(a), (b), (c) and (d) respectively for the results of simulations.
See Fig.~\ref{fig:nested fat shock}(a) (resp. (b)) for an illustration of the  nested fat shock pinned on the left (resp. right) in the top row and the result of a simulation for $r=2$ and $\phase 2$ (resp. $\phase{\neg 2}$) in the bottom row. The hyperlinked video \href{https://drive.google.com/open?id=0B93XrncWWJUtb05WNExMbW94T00}{here} shows the spatiotemporal evolution of the same fat shock.

\subsubsection*{Boundary of the $\phase{j}-\phase{\neg j}$ region}
\label{sec:j-bar j}

In the $\phase{j}-\phase{\neg j}$ boundary,  particles of species $\neg{j-1},\dots,0,\dots,j-1$ participate in the nested fat shock. The velocity of the shock fronts are now zero. Therefore, all these fronts perform a lockstep symmetric random walk in the bulk of the system. All subinterval widths will 
remain constant until the nested fat shock hits the boundary.
When one of the extreme fronts gets pinned to the boundary, the widths of the subinterval containing $j-1$'s and $\neg{j-1}$'s can change, but the sum of their widths will remain constant. The other subinterval widths will remain the same.
While this front is pinned, the other fronts continue to move synchronously until either another one gets pinned or the one stuck to the boundary gets unpinned. If one more (either $j-2$ or $\neg{j-2}$) gets pinned, the same phenomenon will repeat for that species. Note that, for instance, if the nested fat shock gets (temporarily) pinned to the right and then the next shock front also gets pinned, the density of $j-1$'s becomes zero.
More and more shock fronts can get pinned until the fat shock containing $0$'s touches the boundary, at which point the latest front to get pinned can only get unpinned.
We thus end up with an instantaneous profile whose schematic is given in the top row of Fig.~\ref{fig:nested fat shock}(c).
A simulation of the movement of the shock fronts can be seen in the spatiotemporal plot in the bottom row of Fig.~\ref{fig:nested fat shock}(c).
The hyperlinked video \href{https://drive.google.com/open?id=0B93XrncWWJUtbWNYWXNDbnl1bWs}{here} shows the spatiotemporal evolution of the same fat shock.
The steady state density profile can be obtained by averaging over the uniform shock locations and gives rise to piecewise linear profiles for species $\neg{j},\dots,0,\dots,j$. The calculation of these profiles is not difficult, but is tedious and is skipped. The currents have the same values as those in regions $\phase j$ and $\phase{\neg j}$ with $a = b$.

\section*{Discussion}

In this work, we have found the complete phase diagram for a very general multispecies exclusion process, called the mASEP, in contact with reservoirs. We have found two nonequilibrium phenomena, namely dynamical localisation and dynamical exclusion.
Just like the fat shock construction explained the gross features of the semipermeable TASEP in all phases, we find a new object called a nested fat shock which explains the features here. Since the widths of the subintervals in the nested shock are fixed because of the conservation of the total number of $j$ and $\neg j$ species, the system is extremely constrained. These constraints play a crucial role in establishing the structure in the various phases.
It is an interesting open question as to how the phase diagram will look like with more general boundary rates.

The proofs of our results have appealed in a fundamental way to the colouring argument. It is natural to ask how general this argument is. We are working on 
a large class of multispecies exclusion processes by systematically exploiting this argument, and we expect new kinds of dynamical structures to appear.

In recent times, multispecies exclusion processes have found applications in physical, chemical and biological systems, as mentioned in the introduction. It would be interesting to see whether experimental realizations can be found for the various phases that we have shown in this work.

\section*{Acknowledgements}
This work is partially supported by UGC Centre for Advanced Studies.
We thank R. Pandit for many useful discussions.

\end{document}